%% file: ms.tex
\documentclass{emulateapj}
\usepackage{colordvi}
\usepackage{epsfig}
\usepackage{graphicx}
\usepackage{amssymb}
\newcommand{\agew}{t$_{w}$}
\newcommand{\Ha}{\hbox{{\rm H}\kern 0.1em$\alpha$}}
\newcommand{\Hb}{\hbox{{\rm H}\kern 0.1em$\beta$}}
\newcommand{\Hd}{\hbox{{\rm H}\kern 0.1em$\delta$}}
\newcommand{\Hg}{\hbox{{\rm H}\kern 0.1em$\gamma$}}
\newcommand{\Hda}{\hbox{{\rm H}\kern 0.1em$\delta_{\rm{A}}$}}

\newcommand{\MgII}{\hbox{{\rm Mg}\kern 0.1em{\sc ii}}}
\newcommand{\CIV}{\hbox{{\rm C}\kern 0.1em{\sc iv}}}
\newcommand{\NeV}{\hbox{[{\rm Ne}\kern 0.1em{\sc v}]}}
\newcommand{\OII}{\hbox{[{\rm O}\kern 0.1em{\sc ii}]}}
\newcommand{\NeIII}{\hbox{[{\rm Ne}\kern 0.1em{\sc iii}]}}
\newcommand{\OIII}{\hbox{[{\rm O}\kern 0.1em{\sc iii}]}}
\newcommand{\NII}{\hbox{[{\rm N}\kern 0.1em{\sc ii}]}}
\newcommand{\SII}{\hbox{[{\rm S}\kern 0.1em{\sc ii}]}}

\newcommand{\suny}{M$_{\odot}$~y$^{-1}$}
\newcommand{\lssfr}{log(sSFR/Gyr$^{-1}$)}
\newcommand{\lmass}{log(M/M$_{\odot}$)}
\newcommand{\lmassd}{log(M$_{\rm{dyn}}$/M$_{\odot}$)}
 
\newcommand{\mstar}{M$_{\star}$}

\newcommand{\vrot}{$v_{\phi}$}

\newcommand{\siglos}{$\sigma_{\!_{\rm LOS}}$}
\newcommand{\siggas}{$\sigma^{\rm{gas}}_{\!_{\rm LOS}}$}
\newcommand{\sigstar}{$\sigma^{\star}_{\!_{\rm LOS}}$}
\newcommand{\ratio}{$\sigma^{\star}_{\!_{\rm LOS}}$/$\sigma^{\rm{gas}}_{\!_{\rm LOS}}$}
\newcommand{\voversig}{$v_{\phi}/\sigma_{r}$}

\begin{document}

\title{Caught in the act: gas and stellar velocity dispersions in a
  fast quenching compact star-forming galaxy at $\lowercase{z}\sim1.7$}

\author{Guillermo Barro\altaffilmark{1},
Sandra M. Faber\altaffilmark{1},
Avishai Dekel\altaffilmark{2},
Camilla Pacifici\altaffilmark{3},
Pablo G.~P\'{e}rez-Gonz\'{a}lez\altaffilmark{4},
Elisa Toloba\altaffilmark{1},
David C.~Koo\altaffilmark{1}, 
Jonathan R.~Trump\altaffilmark{5,6},
Shigeki Inoue\altaffilmark{2},
Yicheng Guo\altaffilmark{1},
Fengshan Liu\altaffilmark{7},
Joel R. Primack\altaffilmark{8},
Anton M. Koekemoer\altaffilmark{9},
Gabriel Brammer\altaffilmark{9},
Antonio Cava\altaffilmark{10},
Nicolas Cardiel\altaffilmark{4},
Daniel Ceverino\altaffilmark{11},
Carmen Eliche\altaffilmark{4},
Jerome J. Fang\altaffilmark{1}, 
Steven L. Finkelstein\altaffilmark{12},
Dale D.~Kocevski\altaffilmark{13}, 
Rachael C. Livermore\altaffilmark{12},
Elizabeth McGrath\altaffilmark{13}}

\altaffiltext{1}{University of California, Santa Cruz}
\altaffiltext{2}{The Hebrew University}
\altaffiltext{3}{Yonsei University Observatory}
\altaffiltext{4}{Universidad Complutense de Madrid}
\altaffiltext{5}{Pennsylvania State University}
\altaffiltext{6}{Hubble Fellow}
\altaffiltext{7}{Shenyang Normal University}
\altaffiltext{8}{Santa Cruz Institute for Particle Physics}
\altaffiltext{9}{Space Telescope Science Institute}
\altaffiltext{10}{Observatoire de Geneve}
\altaffiltext{11}{Centro de Astrobiologia, CSIC-INTA}
\altaffiltext{12}{The University of Texas at Austin}
\altaffiltext{13}{Colby College}

\slugcomment{Submitted to the Astrophysical Journal} 
%\slugcomment{Last edited: \today}
%\date{Submitted: \today}
%\pagerange{\pageref{firstpage}--\pageref{lastpage}} \pubyear{2008}
%\maketitle
\label{firstpage}
\begin{abstract}  

  We present Keck-I MOSFIRE spectroscopy in the Y and H bands of
  GDN-8231, a massive, compact, star-forming galaxy (SFG) at a
  redshift $z\sim1.7$. Its spectrum reveals both \Ha~and \NII~emission
  lines and strong Balmer absorption lines. The \Ha~and {\it Spitzer}
  MIPS~24~$\mu$m fluxes are both weak, thus indicating a low star
  formation rate of SFR~$\lesssim5-10$~\suny.  This, added to a
  relatively young age of $\sim700$~Myr measured from the absorption
  lines, provides the first direct evidence for a distant galaxy being
  caught in the act of rapidly shutting down its star formation. Such
  quenching allows GDN-8231 to become a compact, quiescent galaxy,
  similar to 3 other galaxies in our sample, by $z\sim1.5$. Moreover,
  the color profile of GDN-8231 shows a bluer center, consistent with
  the predictions of recent simulations for an early phase of
  inside-out quenching. Its line-of-sight velocity dispersion for the
  gas, \siggas$=127\pm32$~km s$^{-1}$, is nearly 40\% smaller than
  that of its stars, \sigstar$=215\pm35$~km s$^{-1}$. High-resolution
  hydro-simulations of galaxies explain such apparently colder gas
  kinematics of up to a factor of $\sim1.5$ with rotating disks being
  viewed at different inclinations and/or centrally concentrated
  star-forming regions. A clear prediction is that their compact,
  quiescent descendants preserve some remnant rotation from their
  star-forming progenitors.
\end{abstract}

\section{Introduction}\label{intro}

The formation and structural evolution of the first quiescent galaxies
at $z\gtrsim2$ has been the subject of considerable discussion in
recent years. Since the first papers reporting their remarkably
compact nature compared to quiescent galaxies of similar stellar mass
at low redshift, many studies have verified their small sizes and
characterized its growth with cosmic time (\citealt{daddi05};
\citealt{trujillo07}; \citealt{buitrago08}; \citealt{cimatti08};
\citealt{dokku08}; \citealt{damjanov09}; \citealt{cassata11};
\citealt{szo12}; \citealt{newman12}; \citealt{carollo13};
\citealt{vdw14}). However, the mechanisms responsible for the
formation of such compact objects are still unclear.

A crucial step forward in understanding these formation processes is
the recent discovery of a population of massive (\lmass$>10$), compact
star-forming galaxies at $z\gtrsim2$ (\citealt{wuyts11b};
\citealt{barro13, barro14}; \citealt{patel13};
\citealt{stefanon13}). These galaxies have similar structural
properties to the compact quiescent population, i.e., spheroidal
morphologies, centrally concentrated luminosity profiles and high
S\'ersic indices. The similar properties and number densities suggest
that these compact star-forming galaxies (SFGs) are the immediate
progenitors of the similarly compact first quiescent galaxies
\citep{barro13}. Compact SFGs are typically found in a dusty
star-forming phase characterized by bright far-IR and sub-mm
detections, and relatively {\it normal} star formation rates (SFRs)
similar to those of other star-forming galaxies of the same mass and
redshift, in what is usually referred to as the star-forming main
sequence (\citealt{mainseq}; \citealt{elbaz07}; \citealt{salim07};
\citealt{pannella09}; \citealt{magdis10}; \citealt{wuyts11a};
\citealt{elbaz11}; \citealt{rodi10b}; \citealt{whitaker12b};
\citealt{pannella14}). However, they have radically different
morphologies suggesting that their compact nature is the result of
strongly dissipative transformation processes, such as mergers
(\citealt{hopkins06}; \citealt{naab07}; \citealt{wuyts10};
\citealt{wellons14}) or accretion-driven disk instabilities
(\citealt{dekel09a}; \citealt{ceverino10}; \citealt{dekel13b};
\citealt{zolotov14}) that funnel a large fraction of their gas
reservoirs into the center, rapidly building up a dense stellar core.

Additional evidence in support of the evolutionary connection between
compact SFGs and quiescent galaxies came recently when NIR
spectroscopy of a sample of compact SFGs revealed emission line widths
of $\sigma=200-300$~km~s$^{-1}$ (\citealt{barro14b};
\citealt{nelson14}), in good agreement with the observed stellar
velocity dispersions of compact quiescent galaxies of similar stellar
mass (\citealt{newman10}; \citealt{vandesande13};
\citealt{bezanson13}; \citealt{belli14b}). However, these measurements
are based on different dynamical tracers, and measured on disjoint
populations. Therefore the implied evolutionary connection between
them is indirect. Some caveats to this evolutionary sequence are, for
example: 1) the broad emission lines may be driven by shocks and
outflows rather than the gravitational potential (\citealt{snewman12};
\citealt{genzel14}) , and 2) if compact SFGs do not quench
immediately, their current dynamics may have little to do with their
eventual transition into quiescent galaxies.

A way forward to address these issues is to study the properties of
both the gas and the stars on the same galaxies. These kinematic
properties can be used to test whether the gas and the stars are
probing the same gravitational potential, and the simultaneous
measurement of emission and absorption lines can be used to estimate
the age and star-formation history (SFH) of their stellar
populations. However, these kind of measurements, are still
observationally challenging, particularly for galaxies at
$z\gtrsim1.5$. Firstly, absorption line measurements (e.g., the Balmer
or the Ca$HK$ lines) require NIR spectrographs and long ($>8$~hr)
integrations on $8-10$~m class telescopes, and probing the stronger
emission lines (e.g., \Ha$\lambda6563$~\AA, \NII$\lambda6584$~\AA,
\OIII$\lambda5007$~\AA, \Hb$\lambda4861$~\AA) on the same galaxies
often requires additional observations in other bands. Secondly, the
spectra of galaxies are typically dominated by either emission or
absorption lines, and the phases in which both sets of lines are
strong are usually short and depend on the SFH of the galaxy.

Interestingly, several papers have identified a substantial population
of recently quenched (i.e., young) compact quiescent galaxies at
$z\gtrsim1.5$ which are the most promising candidates to have both
absorption lines from an old, underlying population, and emission
lines from weak, residual star-formation (\citealt{newman10};
\citealt{dokkum10b}; \citealt{toft12}; \citealt{onodera12};
\citealt{vandesande13}; \citealt{bezanson13}; \citealt{bedregal13};
\citealt{belli14a,belli14b}). Spectroscopic follow-up of these
galaxies have revealed prominent Balmer absorption lines and,
occasionally, emission in \OII. The \OII$\lambda3726+3729$ doublet,
however, is typically unresolved at the resolution of these
observations, and thus it is not a reliable kinematic
indicator. Additional observations of these galaxies for other
emission lines in the \Ha~or \OIII~ranges have so far been
unsuccessful.

In this paper, we present $Y$ and $H$ band spectroscopy of a massive
(\lmass~$=10.75$), compact SFG at $z=1.674$ obtained with the MOSFIRE
multi-object infrared spectrograph (\citealt{mosfire1};
\citealt{mosfire2}) on the Keck-I telescope.  The spectra of GDN-8231
show emission and absorption lines which suggest that the galaxy is
quenching rapidly. We model these lines to estimate the kinematics of
the gas and the stars. In addition, we combine the spectra with
optical medium-band photometry from the SHARDS survey
(\citealt{shards}) and Hubble Space Telescope ({\it HST}) WFC3 grism
spectroscopy in G102 (GO13420; PI=Barro) and G141 (GO11600; PI=Weiner)
to obtain detailed spectral energy distributions (SEDs) for the
quenching galaxy and 3 older quiescent galaxies, observed as a part of
the same MOSFIRE survey. From the modeling of their SEDs, we estimate
their stellar ages and SFHs, and analyze the implications for the
evolutionary path from compact SFG to compact quiescent galaxy.

Throughout the paper, we adopt a flat cosmology with $\Omega_{M}$=0.3,
$\Omega_{\Lambda}$=0.7 and H$_{0}=70$~km~s$^{-1}$~Mpc$^{-1}$ and we
quote magnitudes in the AB system.

\begin{figure*}[t]
\includegraphics[width=7cm,angle=0.]{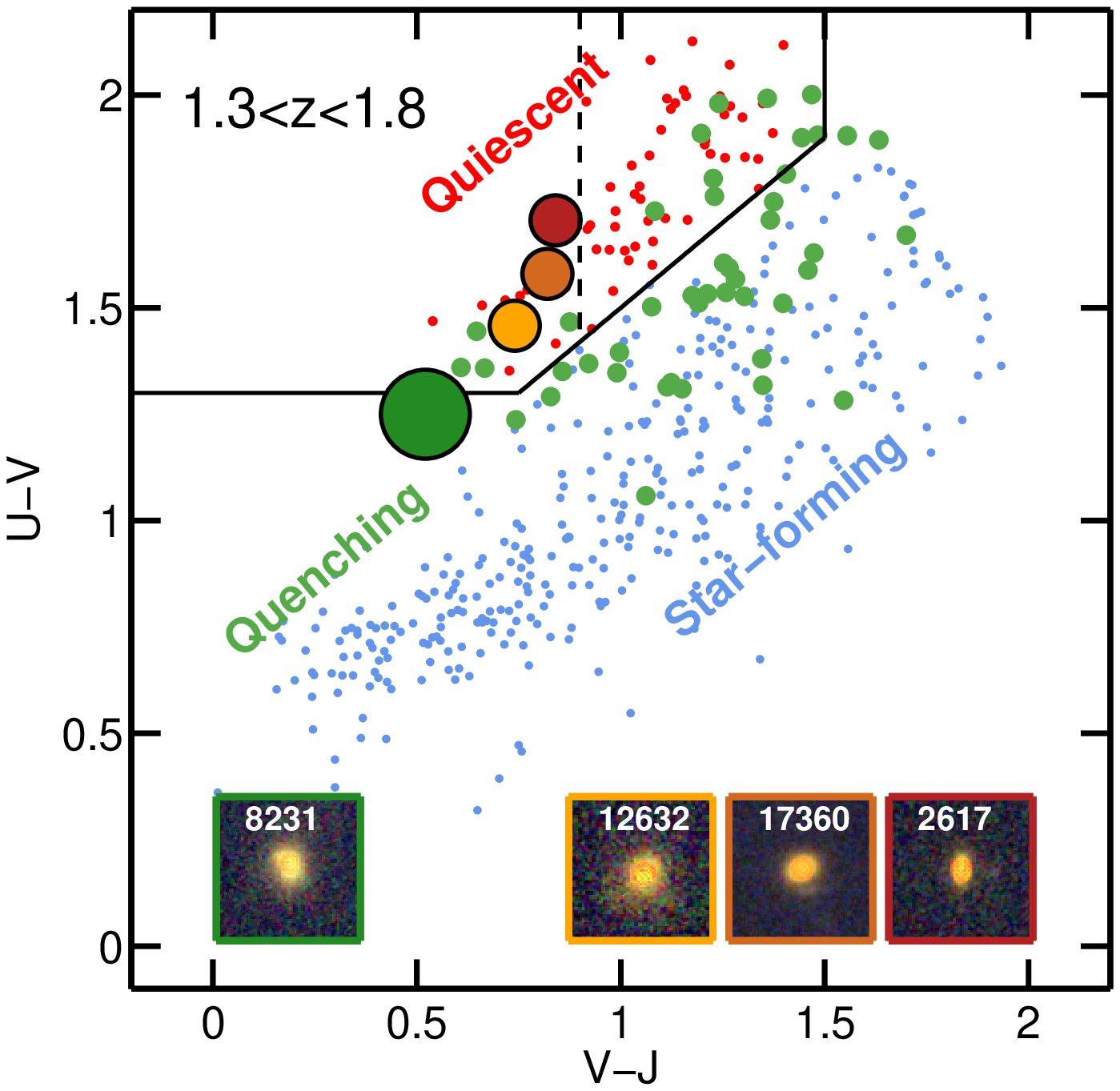}
\includegraphics[width=10cm,angle=0.]{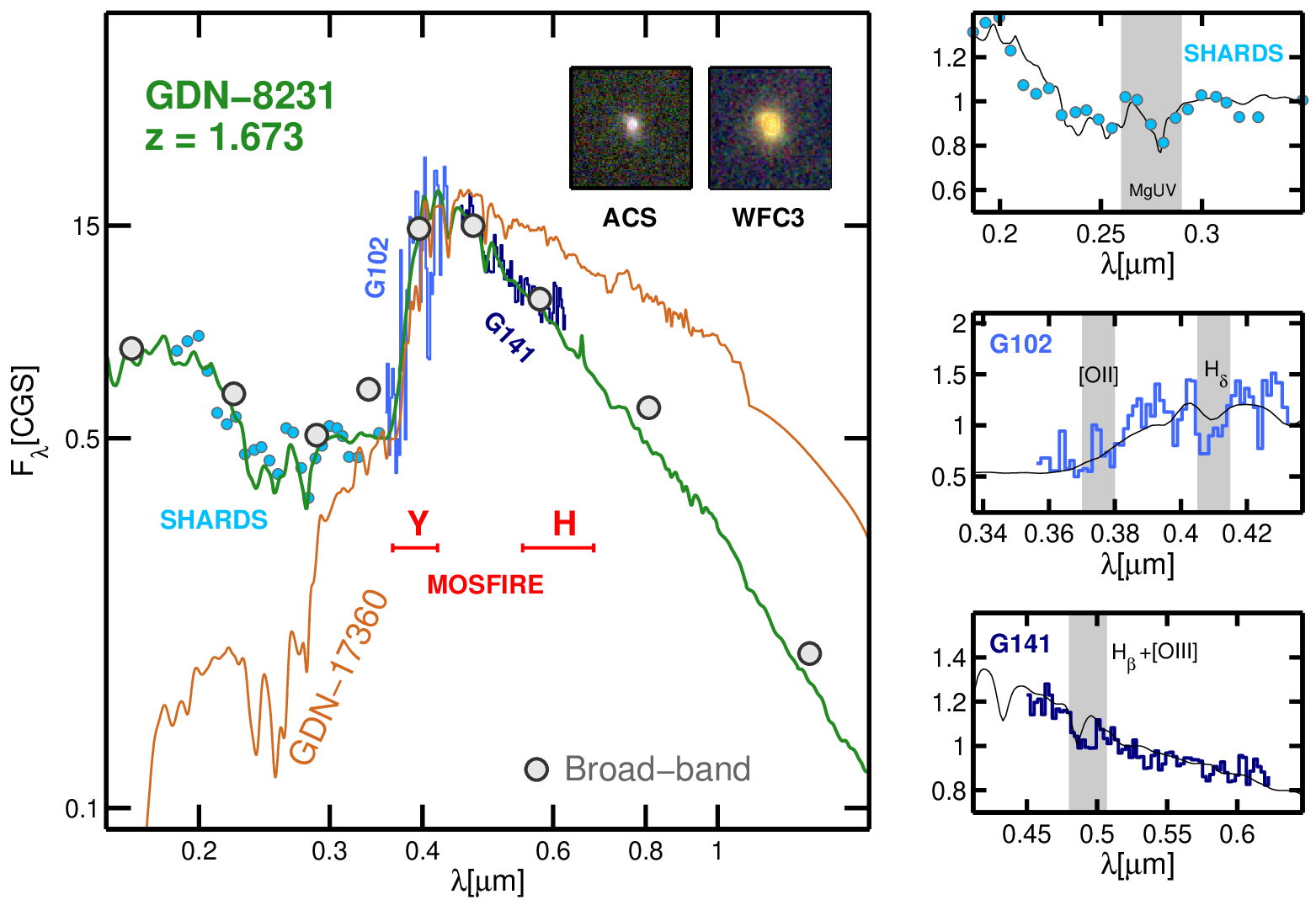}
\caption{\label{selection}{\it Left panel:} UVJ diagram for galaxies
  at $1.3<z<1.8$ more massive than \lmass$>10$ in the GOODS-N
  field. The colors highlight different populations of star-forming
  galaxies (blue; \lssfr$~>-0.25$), quiescent galaxies (red;
  \lssfr~$<-0.75$) and quenching, transition, galaxies (green;
  $-0.25<$~\lssfr~$<-0.75$) identified according to their UV+IR
  SFRs. The large markers show the quenching, compact SFG (GDN-8231;
  dark green) and the 3 quenched galaxies (GDN-2617, 17360, 12632;
  orange to red) observed with MOSFIRE. The {\it HST} color images
  ($zJH$) of the 4 galaxies are shown at the bottom. The location of
  GDN-8231 in the UVJ is consistent with the observed spectral and
  photometric properties, indicating that it is a weakly star-forming
  galaxy transitioning to a quiescent phase.  The quiescent galaxies
  fall within the selection region for recently quenched galaxies
  (left of the dashed line; \citealt{whitaker12}). {\it Right panel:}
  Color images (ACS and WFC3), and composite SED of GDN-8231. The grey
  circles show the (low-resolution) broad-band photometry, the cyan
  markers show the SHARDS medium-band data
  (R$\sim50$;\citealt{shards}) and the light and dark blue lines show
  the {\it HST}/WFC3 G102 and G141 grism spectra. The spectral regions
  probed by $Y$ and $H$ band MOSFIRE spectra are indicated in red. The
  green and orange lines show the best-fit stellar population
  templates from \citet{paci12} at a resolution of R$=50$ for GDN-8231
  and the quiescent galaxy GDN-17360. The 3 sub-panels on the right
  show the zoom-in around the SHARDS, G102 and G141 data highlighting
  spectral features in Mg$_{\rm UV}$, \OII, \Hd~and \Hb.}
\end{figure*} 

\section{Data}
\subsection{Target Selection}\label{data}

We select spectroscopic targets from the CANDELS
(\citealt{candelsgro}; \citealt{candelskoe}) WFC3/F160W ($H$ band)
multi-wavelength catalogs in GOODS-N (Barro et al. in prep.). This
catalog includes deep, UV-to-NIR ground-based observations in several
medium and broad bands, {\it HST} photometry in 9 different bands and
{\it Spitzer}/{\it Herschel} observations in the mid-to-far IR. The
source extraction and merged multi-wavelength spectral energy
distributions (SEDs) are measured following the procedures described
in \citet{galametz13} and \citet{guo13}. In addition, GOODS-N includes
complementary {\it HST}/WFC3 observations in both G102 and G141
grisms, allowing for continuous wavelength coverage from
$0.9<\lambda<1.7$~$\mu$m with a resolution better than R$=130$ (e.g.,
\citealt{3dhst}). The grism data are reduced using the
threedhst\footnotemark~pipeline.  The pipeline handles the combination
and reduction of the dithered exposures, and the extraction of the
individual spectra. These are background subtracted and corrected for
contamination from the overlapping spectra of nearby sources
(\citealt{3dhst}; Momcheva et al. in
prep).\footnotetext{\url{http://code.google.com/p/threedhst}} The
grism data are joined at shorter wavelengths by the 25 optical
medium-bands of the SHARDS survey ($R\sim50$;
\citealt{shards}). Together, these datasets provide remarkable
spectral resolution on a galaxy-by-galaxy basis, that is uniquely
suited for SED-fitting analysis (see \S~3). The stellar masses,
rest-frame colors, and photometric redshifts used for target selection
are derived via SED-fitting using EAzY \citep{eazy} and FAST
\citep{fast}, assuming \cite{bc03} stellar population synthesis
models, a \cite{chabrier} initial mass function (IMF), and the
\cite{calzetti} dust extinction law. The SFRs are computed by adding
the unobscured and obscured star formation, traced by the UV and IR
emission, following the method described in \citet[][see also
  \citealt{barro13,barro14}]{wuyts11a}

We select galaxies in a {\it transition} stage between the
star-forming main sequence and the quiescent, red sequence using
$-0.75 <$\lssfr$<-0.25$. This corresponds to galaxies approximately
$\sim2\sigma$ below the main sequence at $z\sim1.75$
(\citealt{whitaker12b}). For galaxies with \lmass~$>10$, this
threshold roughly corresponds to a SFR~$\gtrsim10$~\suny, which is the
$5\sigma$ detection level with MOSFIRE in $\sim$1h exposures (e.g.,
\citealt{trump13}; \citealt{kriek14}). We also impose the compactness
criterion of \citet{barro13},
$M/r^{1.5}_{\mathrm{e}}=10.3~M_{\odot}$kpc$^{-1.5}$, to select
galaxies with similar structural and morphological properties as the
quiescent population at that redshift. Figure~\ref{selection}
illustrates that the sSFR criterion is consistent with the $U-V$
vs. $V-J$ (hereafter UVJ) rest-frame color criterion that has been
shown to be very successful in identifying quiescent galaxies,
breaking the age/dust degeneracy (\citealt{wuyts07};
\citealt{williams10}; \citealt{whitaker11}). The spread of the {\it
  transition} sample along the wedge of the UVJ quiescent region is
primarily driven by differences in the dust reddening and in the
stellar population ages, which suggests a wide diversity of extinction
levels and SFHs among quenching galaxies (e.g., \citealt{wild14}).

We select the most promising candidates for spectroscopic follow-up by
prioritizing bright galaxies with low dust reddening to maximize the
signal-to-noise (S/N) ratio. This additional restriction
preferentially selects galaxies with bluer UVJ colors, near the
so-called post-starburst region of the UVJ diagram (left of the dashed
line in Figure~\ref{selection}; \citealt{whitaker11};
\citealt{bezanson13}). This region encompasses colors that are typical
of recently quenched galaxies, with low extinction levels and young
stellar ages of $\lesssim1$~Gyr. Galaxies in this region also have, on
average, lower stellar masses (\lmass~$\lesssim10.8$) than older
quiescent galaxies with redder colors (\citealt{newman13b};
\citealt{barro14}).

The redshift is restricted to the range $1.3<z<1.8$ that places the
\Ha~emission line in the $H$ band and several other Balmer lines
around the 4000~\AA~break in the $Y$ band. The observed mask contains
1 candidate that is a {\it transitioning}, compact SFG (GDN-8231;
green circle) and 3 recently quenched galaxies (GDN-12632, GDN-17360
and GDN-2617; orange to red circles). GDN-8231 is relatively massive
(\lmass~$=10.7$) and presents low levels of star-formation ($\rm
SFR=10$~\suny) evidenced by a weak detection in MIPS 24~$\mu$m
($f(24)=40$~$\mu$Jy). The low sSFR and mild extinction ($A_{\rm
  V}=0.3$~mag) are consistent with its location in the UVJ diagram,
and suggests that it is on the brink of becoming a young quiescent
galaxy, similar to the 3 quiescent targets in our sample. The
photometric and spectroscopic galaxy properties of the four galaxies
are summarized in Table~1.

\begin{figure*}[t]
\includegraphics[width=11.3cm, bb=97 47 825 456]{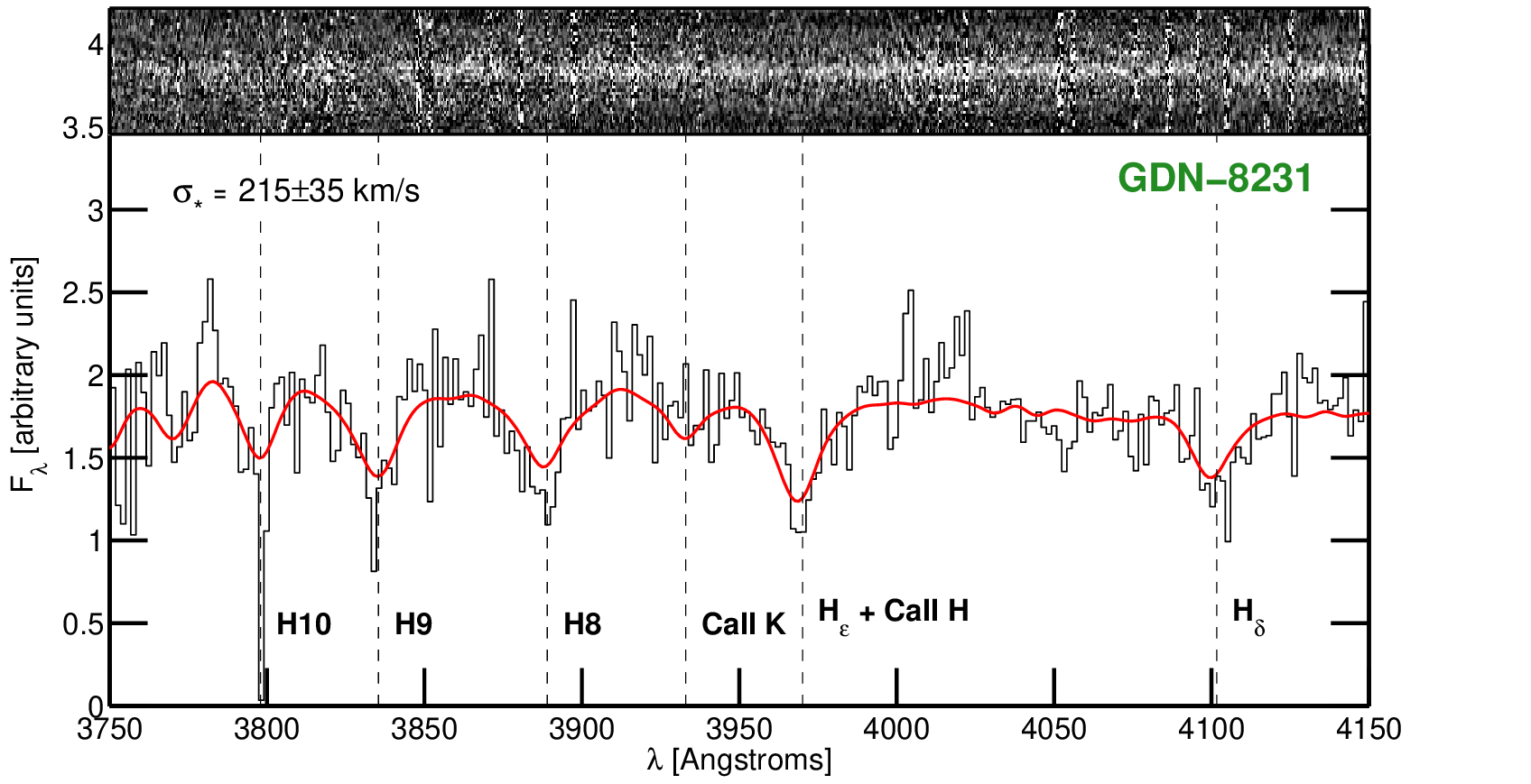}
\includegraphics[width=6.1cm]{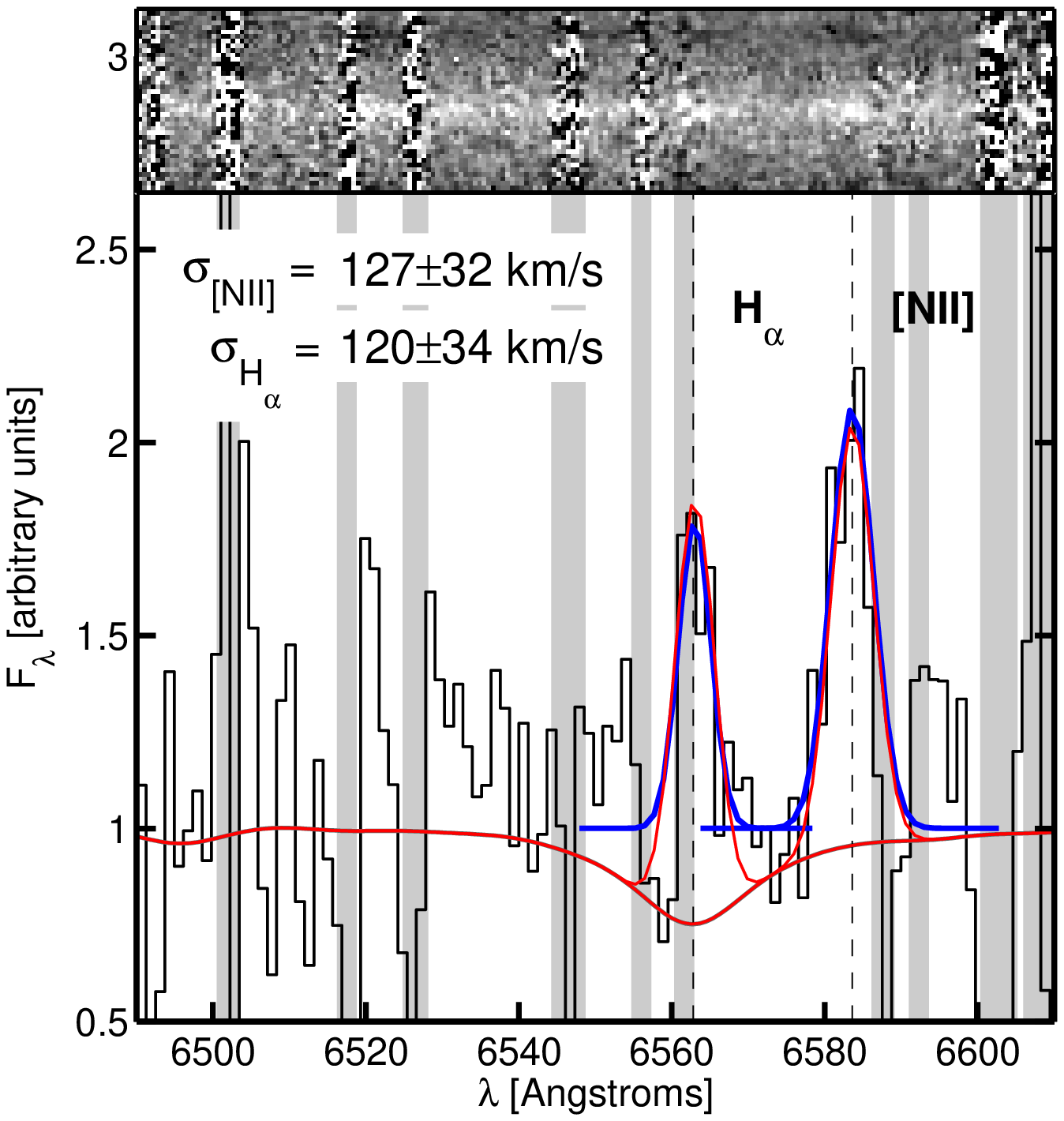}
\caption{\label{spectra}{\it Left panel:} MOSFIRE $Y$ band spectrum of
  the quenching, compact SFG (GDN-8231) with detected absorption
  lines. The best-fit model used to determine the stellar velocity
  dispersion and the absorption indices is shown in red. {\it Right
    panel:} MOSFIRE $H$ band spectrum of GDN-8231 showing the spectral
  range around the \Ha~and \NII~lines. The regions contaminated by sky
  lines are shown in gray. The blue Gaussians show the best fit to the
  \Ha~(\siglos$=97$~km s$^{-1}$) and \NII~(\siglos$=127$~km s$^{-1}$)
  lines using the average continuum level. The red Gaussians show the
  best-fit to the \Ha~line (\siglos$=120$~km s$^{-1}$) correcting for
  the \Ha~stellar absorption as determined from the fit to the $Y$
  band spectrum.}
\end{figure*} 

\subsection{Spectroscopic data}

Data were collected on the nights of 2014 April 17 and May 11 using
the MOSFIRE instrument (\citealt{mosfire1,mosfire2}) on the Keck-I
telescope. The sky conditions were clear and the median seeing ranged
from $0\farcs5-0\farcs7$ seeing. We observed 1 mask configuration in
both the $Y$ ($0.97<\lambda<1.12~\mu$m) and $H$ bands
($1.46<\lambda<1.81~\mu$m), with individual exposure times of 180~s
and 120~s, for a total 5.5~h and 2~h, respectively. We used 2-point
dithers separated by $1\farcs5$ and slit widths of $0\farcs7$. The
instrumental resolution of MOSFIRE with $0\farcs7$ slit widths is
approximately $R=3200$ ($\sim$5~\AA~per resolution element). The 2D
spectra were reduced, sky subtracted, wavelength calibrated, and
one-dimensionally extracted using the public MOSFIRE data reduction
pipeline\footnotemark.

\footnotetext{\url{http://www2.keck.hawaii.edu/inst/mosfire/drp.html}}

\section{Analysis}
\subsection{Kinematic measurements and line ratios}\label{kinmes}

The 4 galaxies are NIR bright ($H<22$~mag) and present clear continuum
detections (S/N$>5$) in both the $Y$ and $H$ band spectra. GDN-8231
exhibits multiple Balmer absorption lines in the $Y$ band spectrum,
from \Hd~to H10, and it is the only galaxy showing \Ha~and
\NII~emission lines in the $H$ band spectrum
(Figure~\ref{spectra}). The other 3 quiescent galaxies also have clear
Balmer absorption lines in the $Y$ band but present no emission lines
in the $H$ band.

We measure the line-of-sight (LOS) stellar velocity dispersion using
the penalized pixel-fitting software pPXF \citep{ppxf}. This software
fits the galaxy spectrum with a model created as a linear combination
of the stellar templates that best reproduce the galaxy spectrum
allowing different weights for each templates. The stellar templates
used are the stars from the MILES stellar library (\citealt{miles};
\citealt{cenarromiles}). Before fitting the galaxy spectrum, we mask
the regions contaminated by telluric atmospheric bands and pixels with
strong residuals from the sky lines, and convolve it with a Gaussian
function whose width is the quadratic difference between the
resolution of the MILES stellar library (FWHM~$=2.5$~\AA;
\citealt{falconmiles}) and the instrumental resolution of our
observations measured in sky lines (FWHM~$=2.7$~\AA). We estimate the
uncertainty in the stellar velocity dispersion by running 1000 Monte
Carlo simulations. In each simulation, the flux of the spectrum is
perturbed within a Gaussian function whose width is the uncertainty in
the flux obtained in the reduction process. The stellar velocity
dispersion is measured in each simulation and the uncertainty is the
standard deviation of the Gaussian distribution of all 1000 velocity
dispersion estimates. Based on this method, we obtain a stellar
velocity dispersion for GDN-8231 of \sigstar$=215\pm35$~km s$^{-1}$
(see Table~1 for the quiescent galaxies).

We measure the LOS gas velocity dispersion, \siggas, of GDN-8231 by
fitting a Gaussian profile to the emission lines, measuring its FWHM,
and subtracting the instrumental broadening in quadrature from the
FWHM. The velocity dispersion is then the corrected FWHM divided by
2.355. We fit the \Ha~and \NII~lines independently. As shown in
Figure~\ref{spectra}, \NII~is detected at higher S/N ratio because the
\Ha~line is partially contaminated by a skyline, and it appears to be
self-absorbed (i.e., the continuum emission is affected by Balmer
absorption). As a result, if we fit the lines using the same continuum
level the velocity dispersion inferred from \Ha~is smaller than that
from \NII, \siggas$=90\pm18$~km s$^{-1}$ and $127\pm32$~km s$^{-1}$,
respectively. Although it is possible for Balmer and forbidden lines
to have different widths if they originate in different regions, the
most likely explanation for such a large difference is the
self-absorption in \Ha. To account for that effect, we use the
best-fit stellar template to the absorption spectra to establish the
continuum level for the \Ha~emission, and we recompute the fit. With
this method, the inferred velocity dispersion is
\siggas~$=120\pm34$~km s$^{-1}$, consistent with the
\NII~result. Although the two measurements agree after accounting for
Balmer absorption, we adopt the higher-S/N \NII~ measurement as the
more reliable tracer of gas velocity dispersion.

The spectra are not flux calibrated. However, we can use the
equivalent width of \Ha~corrected for stellar absorption
(EW(\Ha)$_{\rm corr}=8.3\pm0.7$) and the continuum flux, inferred from
the SED modeling, to calculate the \Ha~line flux and SFR. Using the
empirical relation from \citet{ken98} and the attenuation inferred
from SED-fitting (A$_{\rm V}=0.3$), we obtain values of
SFR~$=3-6$~\suny, depending on the extra nebular extinction with
respect to the continuum (A$_{H\alpha}=2.44-1.86$~A$_{\rm cont}$;
e.g., \citealt{calzetti}; \citealt{price14}). These values are
consistent with the estimate from MIPS~24~$\mu$m data. The \Ha~line
flux corrected for stellar absorption also allows us to estimate the
intrinsic value of the line ratio \NII/\Ha$=1.2$.  Even without
measuring the \OIII/\Hb~line ratio to fully constrain the ionization
diagnostic diagram (i.e., BPT; \citealt{bpt}), an \NII/\Ha~ratio of
the order of unity already suggests that the nebular emission is at
least partially powered by an AGN. The galaxy, however, is not
detected in the 2Ms X-ray data \citep{chandra2m} which implies that
the AGN is relatively weak, perhaps shutting down along with the star
formation in the galaxy. This result, together with the large fraction
of AGNs ($\sim40$\%) found in compact SFGs at $z=2-3$ \citep{barro14},
suggest that the black hole growth at high redshift closely follows
the star-formation history (i.e., it starts and shuts down with the
star-formation), as opposed to the observed trend in the local
Universe, where there seems be to a delay between the starburst and
the peak of the AGN phase (\citealt{wild10}; \citealt{yesuf14};
\citealt{caballero14}). In that case, AGN may play a more active role
in the quenching of star formation at $z\sim2$.

\subsection{Luminosity profile and Morphology}
\label{morph}

\begin{figure}[t]
\centering
\includegraphics[width=8.7cm,angle=0.]{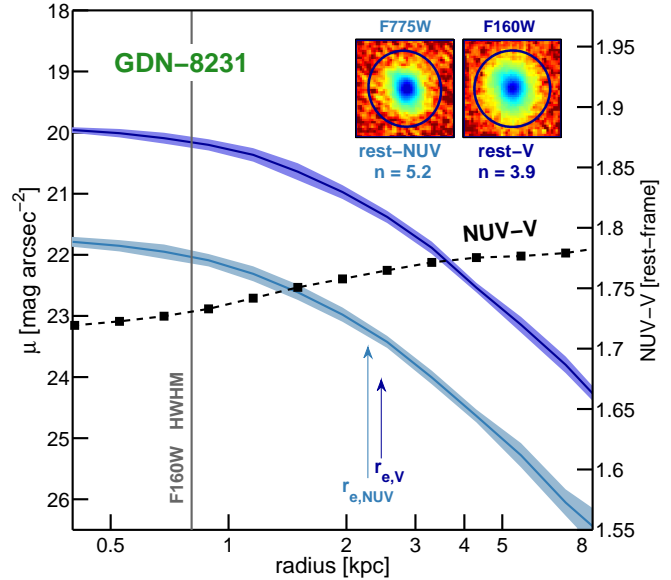}
\caption{\label{profile} Rest-frame surface brightness profile and
  color profile of GDN-8231. The rest-frame profiles are computed by
  interpolating at each radius the best-fit SED derived from the
  observed surface brightness profile in 9 HST bands measured with
  IRAF/ellipse (see \citealt{liu13} and Liu et al. in prep for more
  details).  At $z=1.67$, the rest-frame NUV and V bands roughly
  correspond with the observed $i$ and $H$ bands, respectively. The
  arrows indicate the effective radius in those bands from the
  best-fit S\'ersic profiles obtained using GALFIT. The grey line
  indicates the PSF half width at half-maximum (HWHM) in the $H$
  band. The insets show the images of GDN-8231 in the $i$ (PSF-matched
  to $H$ band) and H bands. The black circle has radius of 1\arcsec
  ($\sim8.4$~kpc). GDN-8231 has an positive color gradient and
  relatively low sSFR, which are consistent with the expectations for
  an early phase of inside-out quenching after a central starburst.}
\end{figure}

The MOSFIRE spectra of GDN-8231 are spatially unresolved and thus do
not provide additional insights on the kinematic profile of the gas
and the stars, or the spatial distribution of star formation. The
high-resolution {\it HST} imaging, however can be used to determine
the overall structural properties and the radial distribution of
star-formation in the galaxy. Figure~\ref{profile} shows the surface
brightness and color profiles of GDN-8231 computed from the fitting of
the HST-based SEDs measured at each radius. The profile shows a
positive color gradient of $d(NUV-V)/dr=6\cdot10^{-3}$~mag/kpc that is
indicated by a bluer center with a 20\% smaller $r_{e}$ in the
rest-frame NUV with respect to the V band ($r_{e,NUV}/r_{e,V}=0.8$).
Assuming that the NUV is a good tracer of the SFR, these measurements
suggest that the SFR profile is more centrally-concentrated than the
stellar profile. A possible interpretation for this result is that we
are witnessing the quenching of a central starburst right after it has
formed a dense stellar core ($\Sigma_{1~\rm
  kpc}=9.7$~M$_{\odot}$~kpc$^{-2}$), which seems to be a pre-requisite
for quiescent galaxies (\citealt{cheung12}; \citealt{fang13};
\citealt{dokkum14}).  Such interpretation is an excellent match for
the predictions of simulations that describe the formation of compact
galaxies as the result of strongly dissipative gas-rich events, such
as mergers and/or disk instabilities (e.g., \citealt{dekel13b};
\citealt{zolotov14}; \citealt{wellons14}). These processes are
characterized by a wet-inflow (i.e., gas-inflow rate $>$ SFR) that
builds-up the central gas density, thereby enhancing the SFR at the
center and growing a dense stellar core.  The weakening of such inflow
marks the onset of inside-out quenching due to gas depletion. In that
onset phase, the SFR is still high and thus the color gradient is
still positive, as observed in GDN-8231. However, as inside-out
quenching progresses, the SFR profile flattens, and the color gradient
turns negative, as observed in compact quiescent galaxies
(\citealt{guo12}; \citealt{wuyts12}; \citealt{szo12}).

GDN-8231 has high-S\'ersic values consistent with other compact
star-forming and quiescent galaxies (\citealt{barro13}). Its visual
appearance in the $H$ band is smooth and spheroidal. However, the ACS
images, that probe the rest-frame UV, show small irregular patches,
perhaps indicative of its prior, dissipative compaction
event. Adopting the assumptions of \citet{miller11}, we infer a
viewing-angle inclination of $i=42^{\circ}$ from an axis-ratio value
of $b/a=0.74$.

\subsection{Spectral indices and SED fitting}\label{indexmeas}

\begin{figure*}[t]
\centering
\includegraphics[width=16cm,angle=0.]{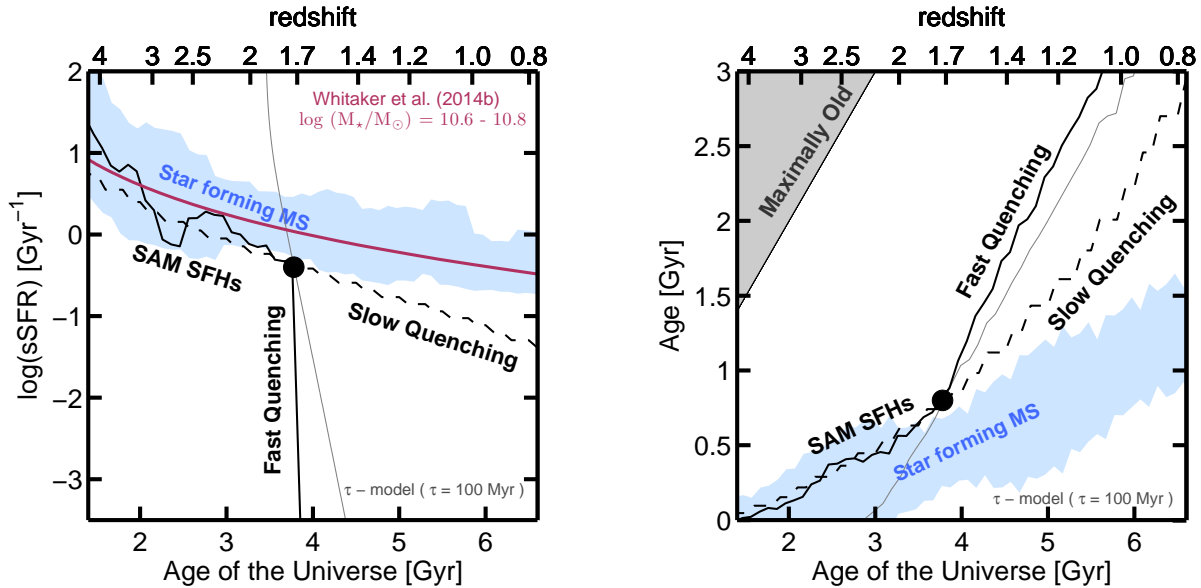}
\caption{\label{cartoon} Evolutionary tracks in sSFR (left) and
  luminosity-weighted age (right) vs. age of the Universe for
  different SFHs. The blue shaded region depicts the star-forming main
  sequence determined from the average SFH of SFGs drawn from the
  model library of \citet{paci12}. This region agrees well with the
  observational results of \citet{whitaker14} for SFGs of intermediate
  mass.  The black lines illustrate the evolution of 2 galaxies that
  have a secular growth from $z\sim4$ to $z=1.7$, followed by either
  fast (solid) or slow (dashed) quenching of star formation. In a fast
  quenching galaxy the luminosity-weighed age grows linearly with time
  (passive evolution). However, in a slow quenching (or main sequence)
  galaxy the luminosity-weighed age increases more slowly (i.e., the
  slope is $<1$).  A $\tau$ model can describe either a fast or slow
  quenching. However, a short $\tau$ (gray line) would provide
  unrealistic results for galaxies described by a two-phase SFH (e.g.,
  a main sequence + fast quenching).}
\end{figure*}

One of the main obstacles to estimating stellar population properties
from the analysis of SEDs is that at the typical resolution of
broad-band surveys (FWHM$\sim$0.2~$\mu$m/(1+z); R$\sim$6) the most
relevant continuum and emission/absorption line features are usually
diluted, which results in large uncertainties in inferred properties
(\citealt{muzzin09}; \citealt{conroygunn}). One way to circumvent this
problem is by using higher spectral resolution data to obtain more
accurate measurements of line strengths and spectral indices, which
are key indicators of stellar age, and present/past star-formation
activity (e.g., \citealt{kelson01}; \citealt{kauffmann03}).

From the MOSFIRE $Y$ band spectra we measure the \Hda~Lick index
\citep{worthey97} using PPXF to estimate the best-fit value and
uncertainty. This index has been traditionally used as an age and SFH
indicator (\citealt{trager00}; \citealt{pat06}). From the $H$ band
spectra we measure the \Ha~Equivalent Width, EW(\Ha), which is
sensitive to the instantaneous (last few Myr) SFR. Only GDN-8231
presents \Ha~emission, the other 3 quiescent galaxies have strong
continuum detections that place reliable upper limits on EW(\Ha). In
addition to the MOSFIRE spectra, we use the SHARDS medium-band data
and the {\it HST} grism data to probe for additional absorption
features and continuum breaks. The right panel of
Figure~\ref{selection} illustrates the high spectral resolution
($\sim10-100\times$ better than broad-band filters) of the merged
medium-band/grism SED, revealing \MgII~absorption at 2800~\AA~in
SHARDS and Balmer absorption lines and the 4000~\AA~break in G102 and
G141. We quantify the later using the D4000 index (e.g.,
\citealt{balogh99}; \citealt{kauffmann03}).  The G102 grism shows also
a weak \OII~emission line (EW~$15\pm5$~\AA).  However, there is no
clear sign of \OIII~emission in G141. The later is expected to have
also low EW~$\lesssim10$~\AA, and thus can be partially hidden by the
\Hb~absorption, as hinted in the stacked spectra of recently quenched
galaxies in \citet{whitaker13}.

The spectral indices are usually analyzed by comparing measurements to
a grid of values derived from stellar population synthesis
models. Here we follow a slightly different approach to combine
information from both the indices and the overall UV-to-NIR SED by
using of the SED-fitting code of \citet[][hereafter P12]{paci12}. The
code performs a simultaneous fitting of the low and high resolution
data and includes priors on the EW of emission and absorption lines to
obtain better constraints on the SFR and SFH of the galaxy (see also
\citealt{paci14,paci15} and \citealt{barro14} for more details). The
galaxy templates are computed from non-parametric SFHs adapted from
semi-analytic models (SAMs) and include both the stellar continuum and
the nebular emission. The SAM SFHs provide a richer parameter space
featuring short-timescale variations of the SFR (burst and
truncations) that are missing in exponentially declining
(exp[-t/$\tau$]) or delayed (t$\times$exp[-t/$\tau$]) models (see also
\S~\ref{stellarpop}).

\section{Results}\label{results}

\subsection{Stellar ages and SFHs}
\label{stellarpop}

\begin{figure*}[t]
\centering
\includegraphics[width=15.cm,angle=0.]{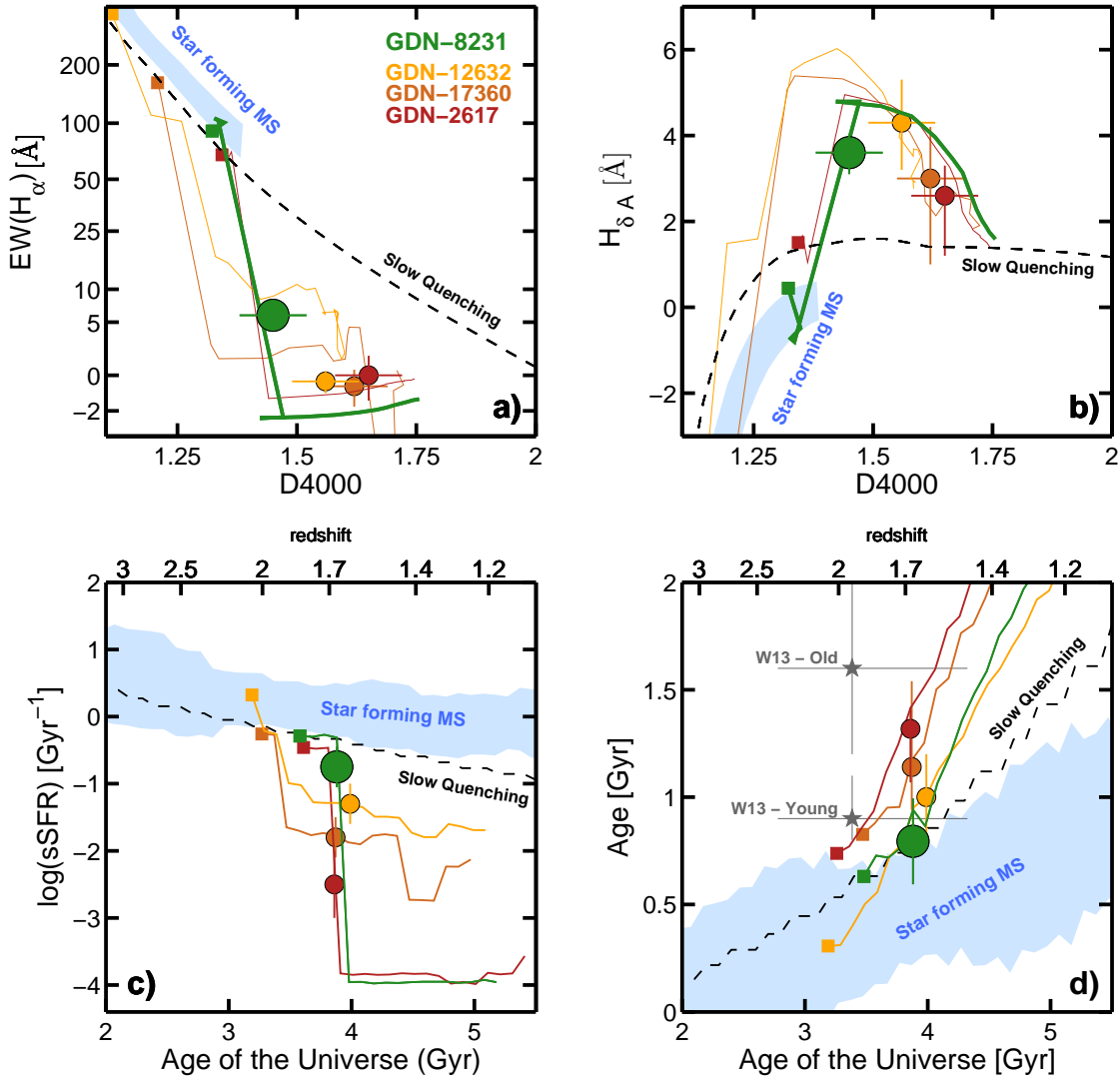}
\caption{\label{indices} {\it Panels a) \& b):} Index-index diagrams
  for the galaxies in the sample. The circles show the values measured
  in the spectra. The colors are the same as in
  Figure~\ref{selection}. The colored lines depict the evolutionary
  tracks for the best-fit SAM SFHs. The tracks start at the onset of
  quenching (colored squares) and continue for $\sim$1~Gyr after the
  redshift of observation. The blue shaded region and the dashed line
  show the evolution of the star-forming main sequence and the slow
  quenching galaxy depicted in Figure~\ref{cartoon} and Panels c) and
  d). The \Ha~and \Hd~indices are sensitive to recent changes in the
  SFR, while D4000 is a good tracer of the SFH averaged over longer
  timescales. In a fast quenching, the EW(\Ha) drops abruptly while
  \Hda~increases due to the appearance of prominent Balmer absorption
  lines. In contrast, a slow quenching galaxy has higher EW(\Ha) and
  weaker \Hda~for the same value of the D4000 index. {\it Panels c) \&
    d):} Same as Figure~\ref{cartoon} for the galaxies in our
  sample. The grey stars show the age of the oldest and youngest
  quiescent galaxies at $z\sim2$ in \citealt{whitaker13}. All 4
  galaxies are consistent with fast quenching. However, the details of
  the quenching path lead to different luminosity-weighted ages. For
  example, GDN-12632 was the first of quiescent galaxies to quench and
  is the youngest, while GDN-2617 was the last to quench and it is the
  oldest.}
\end{figure*}

The stellar age and SFH are not independent properties and thus, the
chosen parametrization of the latter (single burst, N-bursts, $\tau$
models, SAMs, etc) often leads to strong degeneracies in the age and
the characteristic timescale(s) of star formation (e.g., age-$\tau$;
\citealt{conroy13}). These degeneracies, however, can be reduced using
higher spectral-resolution datasets to probe features that are
sensitive to different star-formation and quenching timescales
(\citealt{kriek11}; \citealt{shards}; \citealt{belli15}).

Figure~\ref{cartoon} illustrates the implications of assuming
different SFHs to estimate the age of the galaxies. The blue shaded
region shows the evolution of the sSFR and luminosity-weighted age vs.
time for galaxies in the star-forming main sequence, as described by
the SAM SFHs in the model library of \citet{paci12}. Those galaxies
are thought to be growing in a relatively smooth, secular mode
(\citealt{elbaz07}; \citealt{rodi10b}) in which gas inflow and SFR
have reached a steady-state phase (e.g., \citealt{dekel13a}). The main
sequence pictured by SAM SFHs follows closely the observational
results (e.g., \citealt{whitaker14}) which are also in good agreement
with the predictions of other empirical models (e.g.,
\citealt{peng10}; \citealt{behroozi13}). In the main sequence
paradigm, quenching can be interpreted as the departure from a stable
growth phase, which can be either fast (e.g., SFR truncation; solid
line) or slow (shallower slope; dashed line). After fast quenching,
age increases linearly with time (i.e., passive evolution), while for
slower quenching the slope is shallower, and the age increases slowly
due to slowly declining star-formation (right panel of
Figure~\ref{cartoon}). In $\tau$ models, the timescale controls the
quenching time, and it can be tuned to describe fast or slow quenching
(grey line in Figure~\ref{cartoon}). However, as one parameter
characterizes the whole SFH, the prediction is not realistic for
galaxies that are described by 2 or more distinct phases, such as a
secular growth followed by fast quenching.

Panels a) and b) of Figure~\ref{indices} show index-index diagrams
sensitive to the SFH and quenching time. The colored circles indicate
the values measured in the MOSFIRE spectra of the 4 galaxies, and the
lines show the evolutionary tracks of their best-fit SAM SFHs. The
D4000 index is a good tracer of the average age of the stellar
populations, while the EW(\Ha) and \Hda~are more sensitive to recent
star-formation. For slow quenching, the EW(\Ha) in emission decreases
slowly while the Balmer absorption is relatively small and constant
with time, \Hda~$\sim1.8$. In contrast, fast quenching causes a rapid
decline in the \Ha~emission followed by an absorption plateau with
EW(\Ha)$\sim-2$~\AA, and a strong \Hda~absorption. This relatively
short-lived phase ($\sim$1~Gyr), characterized by the lack of
\Ha~emission and strong absorption in other Balmer lines, typical of
A-stars, is known as a post-starburst phase (pSB; e.g.,
\citealt{wild10}).

The spectral properties of GDN-8231 agree well with the sSFR selection
criteria indicating that the galaxy is on a fast quenching
path. Besides the \Ha~emission, which indicates the presence of weak,
ongoing star-formation, the SED fit, the D4000 index and the
\Hda~values suggest that the emission is rapidly declining. In fact,
given the short quenching time, the chances of finding a galaxy in
this phase are small, which makes GDN-8231 quite unique. The 3
quiescent galaxies in our sample are further along in their evolution,
as indicated by their stronger D4000~$\sim1.6$.  However their indices
and best-fit SFHs are also consistent with fast quenching. The similar
SFHs are another indication that GDN-8231 is evolutionarily linked to
the quenched galaxies. For a definition of quenching time,
$t_{\rm{q}}$, as the elapsed time from having EW(\Ha) values
consistent with those of a main sequence galaxy, to EW(\Ha)$\sim0$
(tracks in panel-a), the values range from a few Myrs for GDN-8231
(i.e., SFR truncation) to 500~Myr for GDN-17360.  Interestingly the
SAM SFHs suggest that galaxies show pSB features as a result of fast
quenching from the star-forming main sequence, without being preceded
by a strong starburst (i.e., $\sim3\times$ higher SFR than the main
sequence). A more practical definition of $t_{\rm{q}}$ as the elapsed
time from \lssfr$=0$ to -1 (tracks in panel c)), results in similar
values of $t_{\rm{q}}=300-800$~Myr.

The fast quenching scenario for all 4 galaxies agrees well with recent
works that find strong Balmer absorptions on similarly young quiescent
galaxies at $z\sim1.5$ (\citealt{newman10}; \citealt{onodera12};
\citealt{bezanson13}; \citealt{vandesande13}; \citealt{belli14a}).  It
differs, however, with the results of \citet{kriek11}, who studied a
sample of stacked, high-resolution SEDs of galaxies at $0.5<z<2.0$
finding EW(\Ha) values consistent with gradually declining
SFHs. Assuming the difference is not caused by the slightly different
redshift range, a plausible explanation for this discrepancy is that
there are indeed different evolutionary quenching tracks (e.g.,
\citealt{martin07}; \citealt{goncal12}; \citealt{schawinski14} or
\citealt{belli15}), and current spectroscopic samples at $z\gtrsim1.5$
are still not fully representative of the whole quenching population.

\begin{figure}[t]
\includegraphics[width=8.5cm,angle=0.]{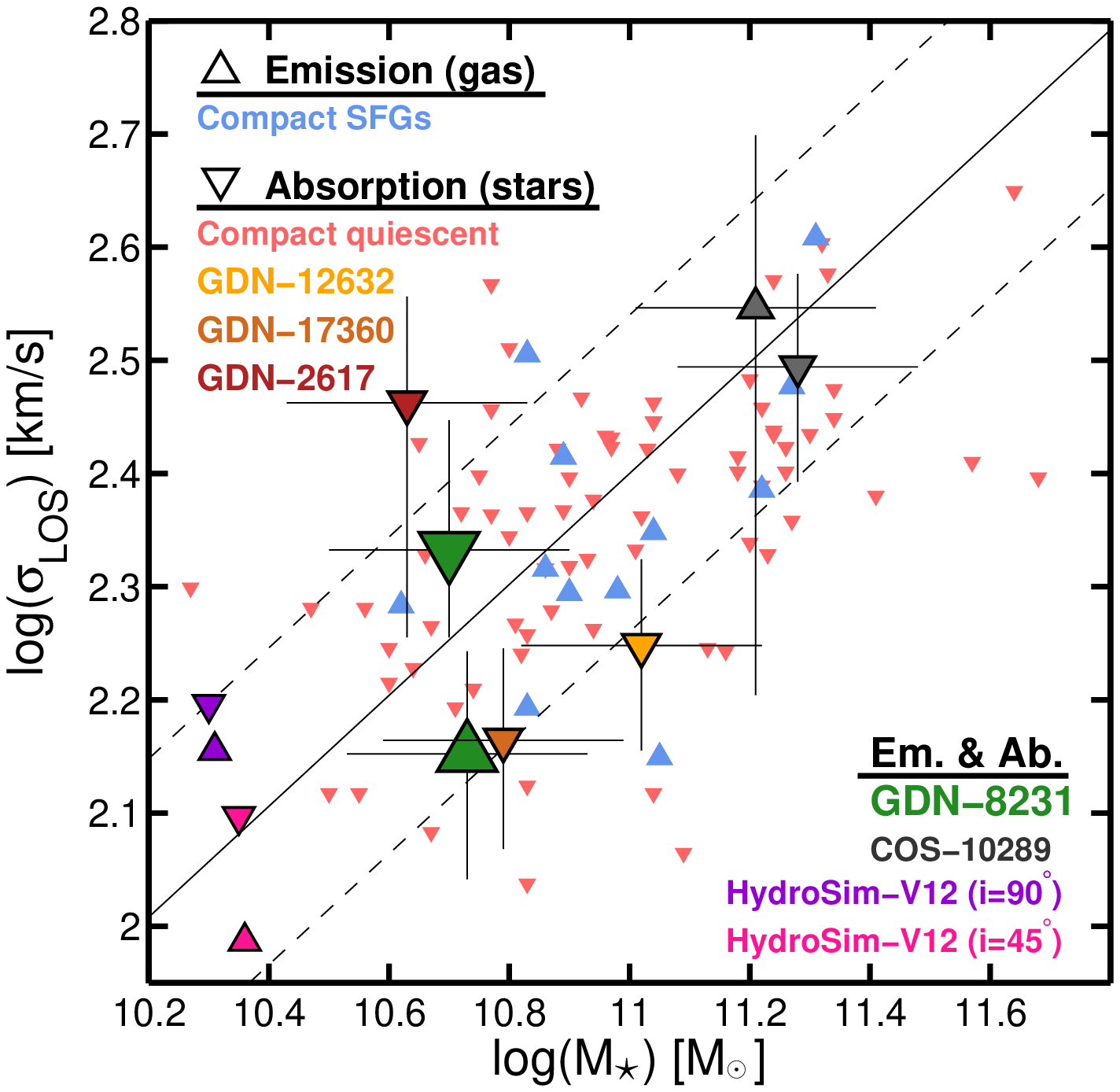}
\caption{\label{massigma} \siglos~vs. \mstar~for different galaxy
  samples. The point up and down triangles show emission (gas) and
  absorption (stars) line measurements, respectively. The blue
  triangles depict 13 compact SFGs in \citet{barro14}, and one galaxy
  from \citet{nelson14}. The red triangles show a compilation of
  quiescent galaxies (\citealt{vandesande13}; \citealt{bezanson13};
  \citealt{belli14a,belli14b}) at $z\gtrsim1.5$. The overlapping
  distributions for compact SFGs and quiescent galaxies suggests that
  both populations have similar kinematic properties. This is
  supported by the agreement in the gas and stellar dispersions of
  COS-10289 (\citealt{belli14b}, and \citealt{barro14b} for the
  stellar and gas kinematics). However, GDN-8231 (green) has
  $\sim40\%$ lower dispersion in the gas than in the stars. A possible
  explanation is that the gas has colder kinematics than the stars
  (\voversig~$>1$), and thus its line-of-sight dispersion is prone to
  stronger projection effects. The gas and stellar dispersion for the
  simulated galaxy V12 (see also Figure~\ref{simul}) illustrates the
  bias towards lower values of \siggas~with respect to \sigstar~for
  viewing-angles closer to face-on (magenta).}
\end{figure}

Panel d) of Figure~\ref{indices} shows that the luminosity-weighted
ages of the sample ranges from \agew~$=700$~Myr for GDN-8231 to
\agew~$=1.1-1.3$~Gyr for the 3 quiescent galaxies. The SAM SFHs also
suggest a formation redshift (i.e., onset of star formation) of
$z_{\rm{form}}\gtrsim6$, a half-mass assembly by $z\sim3$ and the
onset of quenching by $z\sim2$. Interestingly, the latter is similar
to the $z_{\rm{form}}$ of the galaxies inferred from $\tau$ models,
which suggests a rapid build up ($\tau<100$~Myr). This, however, is
most likely a limitation of single-parameter $\tau$ models, which are
biased towards short values of $\tau$ in order to reproduce the strong
pSB features in the SED. Secular SFHs with rise and decay timescales
of several hundred Myr appear to be more realistic, and are predicted
by physically motivated models (e.g., \citealt{peng10};
\citealt{behroozi13}; \citealt{gladders13}). However, obtaining a
direct, reliable measurement of the formation timescale would require
additional measurements such as the metallicity [Z/H] or the $\alpha$
element abundance [$\alpha$/Fe], which trace the evolution of the
metals (e.g., \citealt{conroy13}; \citealt{onodera14}).

\begin{figure*}
\centering
\includegraphics[width=15cm,angle=0.]{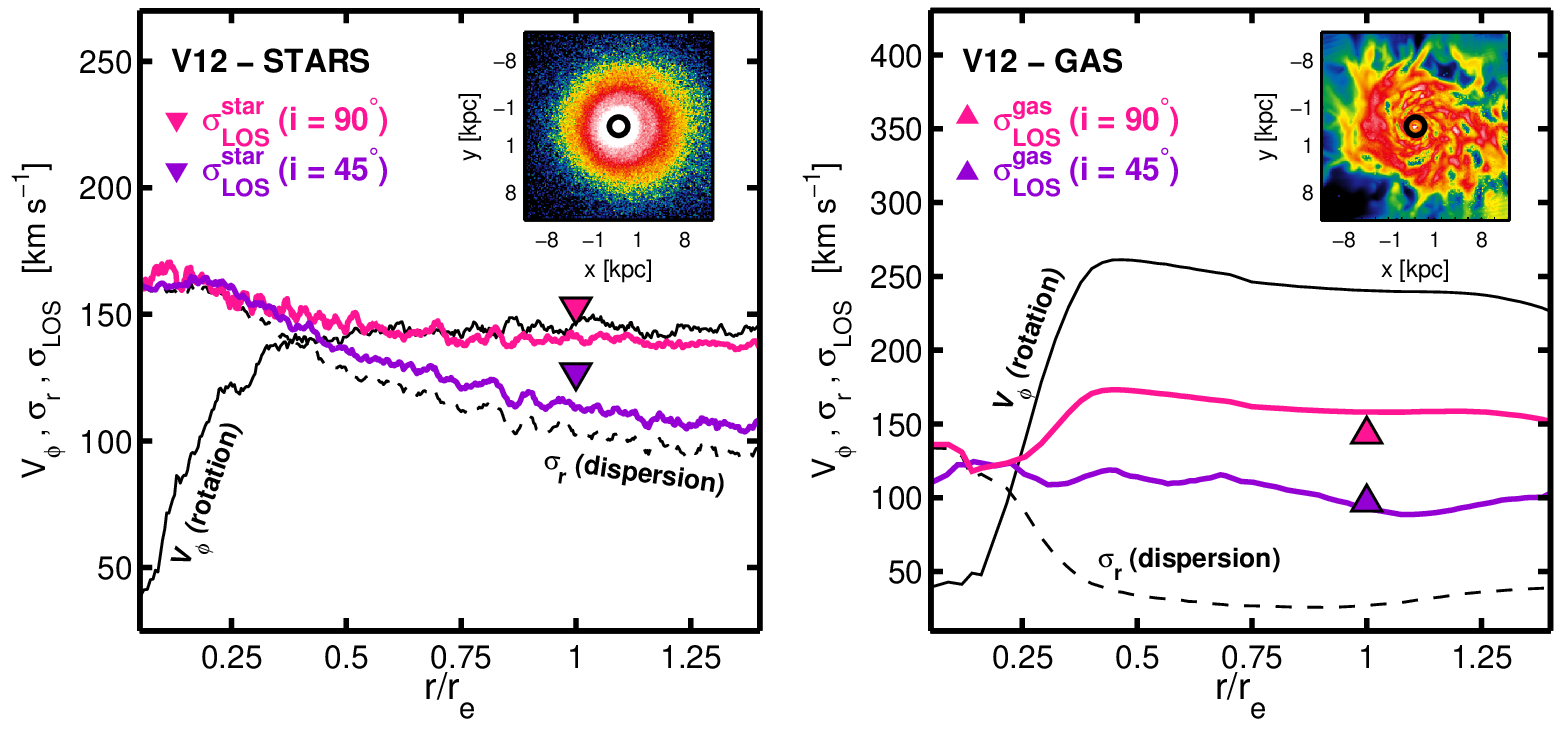}
\caption{\label{simul} Stellar (left) and gas (right) kinematic
  profiles for the simulated galaxy V12 at $z=2.3$ \citep{ceverino14,
    zolotov14}. The kinematic properties are measured in cylindrical
  beams with a depth of 8~kpc. The solid and dashed black lines depict
  the intrinsic rotation and dispersion. The magenta and purple lines
  show the \siglos~profiles for a line-of-sight inclination of
  $i=90^{\circ}$ (edge-on) and $i=45^{\circ}$. The triangles show the
  integrated, mass-weighted values at the $r=r_{\rm e}$. The
  line-of-sight dispersion can be written as $\sigma_{\rm
    LOS}^{2}=\beta(\sin i~v_{\phi})^{2}+\sigma_{r}^{2}$, where $\beta$
  depends on the inclination and density profile of the galaxy. This
  implies that for a rotation dominated component the value of
  \siglos~depends more strongly on projection effects. In V12 the gas
  has higher rotation than the stars (\voversig~$\sim1.4$
  vs. \voversig~$\sim5$) and therefore, the ratio of velocity
  dispersions can be as high as \ratio~$\sim1.5$ for a LOS inclination
  of $i=45^{\circ}$. The insets in the upper-right show the
  10$\times$10~kpc face-on density maps for the stars and the gas. The
  black circle has $r=r_{\rm e}$.}
\end{figure*} 

The stars in panel d) of Figure~\ref{indices} show the age of the
oldest, \agew$=1.5$~Gyr, and youngest, \agew$=0.9$~Gyr, quiescent
galaxies at $z\sim2$ from \citet{whitaker13} (see also
\citealt{newman13b}). The evolutionary tracks of the quiescent
galaxies in our sample are consistent with being direct descendants of
the youngest galaxies of \citet{whitaker13}, and GDN-8231 reaches a
similar age by $z=1.5$. This implies that as new quenching galaxies
join the red sequence the age spread in the quiescent population will
increase by roughly 1~Gyr from $z=2$ to $z=1.5$. Interestingly, if all
the galaxies in our sample follow a {\it pure} passive evolution since
$z=1.7$, their ages by $z=0.6$ would be \agew~$\sim4.5-5$~Gyr, which
is roughly 2~Gyr older than the results of \citet{choi14} for the
oldest quiescent galaxies at that redshift. This could be an
indication that some quiescent galaxies retain low levels of star
formation, or perhaps experience minor wet mergers that rejuvenate
star formation.

In summary, the 4 galaxies in our sample have relatively young ages of
\agew$\lesssim1$~Gyr and present spectral features consistent with
fast quenching of the star formation. In particular, GDN-8231 seems to
be caught in an early stage of a rapid truncation.

\subsection{Kinematic properties}

Figure~\ref{massigma} shows the \mstar$-$\siglos~relation for a
compilation of compact SFGs from \citet{barro14} and quiescent
galaxies at $z\gtrsim1.5$ from the literature (\citealt{vandesande13};
\citealt{bezanson13}; \citealt{belli14a,belli14b}). In
\citet{barro14}, we found that 1) the gas kinematics of massive,
compact SFGs are consistent with the stellar kinematics of compact
quiescent galaxies, and 2) the dynamical masses of both populations
are in good agreement with their stellar masses. We interpreted those
similarities as evidence for an evolutionary connection. In
particular, since both populations have similar structural, kinematic
and dynamical properties, compact SFGs simply turn into compact
quiescent galaxies by quenching their star formation. In addition, one
galaxy (COS-10289), seemed to support this scenario by showing
agreement between its gas and stellar kinematics. However, the low S/N
ratio of the emission line measurement and the presence of an X-ray
AGN within the galaxy place caveats on the interpretation of the
kinematic properties of the gas.

Quite surprisingly, the better-constrained measurements for GDN-8231
(green triangles in Figure~\ref{massigma}) show that the gas has a
lower velocity dispersion than the stars by a factor of
\ratio~$=1.7\pm0.5$. Naively, we expected a value $\sim1$ or, given
that GDN-8231 is quenching, larger values of \siggas~caused by strong
feedback processes (e.g., shocks or outflows; \citealt{stanic12};
\citealt{genzel14}). In turn, the smaller values of \siggas~suggest
that the gas is in dynamical equilibrium and that it has {\it colder}
kinematic properties than the stars (i.e., higher \voversig~in the
gas). If that is indeed the case, the lower dispersion in the gas
could be the result of: 1) widespread star-formation activity in a
disk observed at low inclination (i.e., close to edge-on), or 2) a
centrally concentrated star-forming region probing, on average, lower
values of the rotational velocity \vrot, which grows inside-out. As
described in \S~\ref{morph}, GDN-8231, has an inclination of
$i=42^{\circ}$, and 20\% smaller effective radius in the rest-frame
NUV, which can lead to the smaller values of \siggas~compared to
\sigstar.

In order to provide a better intuition, and quantify how much
projection effects and/or a concentrated SFR profile affect the
measurement of \siglos, we study the kinematic profiles of the gas and
the stars in V12, a high-resolution, hydrodynamic galaxy simulation
($\sim$25~pc grid) drawn from the sample of \citet{zolotov14} and
\citet{ceverino14}. As described in \citet{zolotov14}, these galaxies
have similar stellar and structural evolution as the compact SFGs and
therefore, V12 provides a excellent proxy for the analysis of the
kinematic properties of GDN-8231. Figure~\ref{simul} shows that the
stars in V12 have comparable rotation and dispersion whereas the gas
is rotation dominated (stellar \voversig$\sim1.4$ vs. gas
\voversig$\sim5$). As a result, \siggas~is more sensitive to
projection effects and shows smaller values for low inclinations. For
example, the integrated, mass-weighted \siggas~at $r=r_{\rm e}$ show
ratios of \ratio~$\sim1.0$ and 1.5 for an edge-on ($i=90^{\circ}$) and
an intermediate inclination ($i=45^{\circ}$), respectively (see values
in Figure~\ref{massigma}). Therefore, we conclude that a large
fraction of the observed ratio of velocity dispersions in GDN-8231 can
be accounted for by projection effects.

On the other hand, given the relatively flat \siggas~profile of V12
(right panel of Figure~\ref{simul}), the change in the integrated
value of \siggas($r_{\rm e}$) would be small for a gas density
($\propto$SFR) profile more centrally concentrated than the stellar
profile (i.e., $r_{\rm e,\star}>r_{\rm e,gas}$).  Using the ratio of
effective radii in GDN-8231 ($r_{e,NUV}/r_{e,V}=0.8$) as an example of
different gas-to-stellar concentrations, we find that integrating
\siggas~only up $r=0.8r_{e}$ decreases its value by less than 5\% for
the edge-on case. Nevertheless, the effect of the gas density profile
in \siggas~can be larger in GDN-8231 if: 1) it had a slowly-increasing
rotation curve and a flat $\sigma_{r}$ in the center, and/or 2) the
emission line region, traced by \Ha, had an even smaller $r_{\rm e}$
than the NUV luminosity profile. The latter is more plausible given
the quenching nature of GDN-8231 and the shorter star-formation
timescales probed by \Ha.

Lastly, note that Figure~\ref{simul} shows only the radial component
of the intrinsic dispersion ($\sigma_{\rm r}$) for V12. In the case of
anisotropic dispersion, for example due to strong collimated winds in
the center of the galaxy ($\sigma_{\rm z}\gg\sigma_{\rm r}$), the
observed value of \siggas~for a face-on inclination would be much
larger than that of \sigstar.

\subsubsection{Dynamical mass}

We estimate the dynamical mass of GDN-8231 from \sigstar which, as
described above, is less sensitive to projection effects. Following
the virial equation:
\begin{equation}
M_{\rm{dyn}}(r<r_{e}) = K\frac{\sigma^{2}_{\rm{LOS}}r_{e}}{G}
\end{equation}
where K depends on the galaxy's mass distribution, the inclination,
and velocity field.  We use $K=2.5$, which is the most widely adopted
value for stars in dispersion dominated galaxies (e.g.,
\citealt{newman10}; \citealt{vandesande13}; \citealt{belli14b}), and
is valid under a variety of galaxy geometries and mass distributions
(e.g., \citealt{binney08}). We apply a small correction to scale the
observed \sigstar~to the value in a circular aperture of radius
$r_{e}$, $\sigma_{e}=1.05\times$\sigstar~(\citealt{cappellari06};
\citealt{vandesande11}). The inferred dynamical mass \lmassd~$=11.1$
is slightly larger than the stellar mass, but still consistent within
the uncertainties in both the stellar mass and velocity
dispersion. Previous works have reported a similar offset towards
larger dynamical masses of $\log(M_{\rm dyn}/M_{\star})=0.1-0.16$ for
recently quenched quiescent galaxies $z\gtrsim1.5$
(\citealt{belli14b}).
%and, in fact,
%the 3 quiescent galaxies in our sample also have dynamical masses
%$\sim0.1-0.2$~dex larger than their stellar masses (see Table~1).

\section{Conclusions}

We present Keck-I MOSFIRE NIR spectroscopy of GDN-8231, a massive,
compact SFG galaxy at $z\sim1.7$. This galaxy was selected with sSFR
and rest-frame colors matching an intermediate stage between
star-forming and quiescent. The $Y$ and $H$ band spectra reveal strong
Balmer absorption lines and \Ha~and \NII~in emission. The emission and
absorption lines yield spectral indices and the kinematics of the gas
and the stars.

The spectral indices, SED-modeling, and the comparison to 3 compact
quiescent galaxies at similar redshift, indicate that GDN-8231 was
caught in a rare, early stage of fast quenching. Still relatively
young, with a luminosity-weighted age of $700\pm250$~Myr, GDN-8231
will mature to become a compact quiescent galaxy by redshift
$z=1.5$. The rapid truncation of the SFR is evidenced by the low
EW(\Ha) and weak MIPS~24$\mu$m flux. The color profile is bluer in the
center, which is consistent with the predictions of recent simulations
for an early stage of inside-out quenching. The line ratio of
\NII/\Ha~$\sim1$ suggests the presence of a weak (not X-ray detected)
AGN, a common finding among most compact SFGs (Barro et
al. 2014a). Using SFHs based on SAMs, we find that the assembly of
GDN-8231 is consistent with an early ($z\sim6$) onset of
star-formation, a secular build-up in the star-forming main sequence,
forming 50\% of its stellar mass before $z=3$, and fast quenching at
$z\sim1.7$.

In Barro et al. (2014b), we found that compact SFGs and quiescent
galaxies have similar line-of-sight velocity dispersions for the gas
and the stars, suggesting similar kinematics in progenitors and
descendants. In GDN-8231, however, the dispersion of the gas,
\siggas$=147\pm32$~km~s$^{-1}$, is 40\% smaller than that of the
stars, \sigstar$=215\pm35$~km~s$^{-1}$. This difference can be
explained if the gas has colder kinematics (rotation dominated) than
the stars, and therefore: a) \siggas~is smaller if the viewing-angle
is low (close to face-on), and b) \siggas~is smaller if the
emission-line (star-forming) region is concentrated at the center of
the galaxy, and thus probes low values of \vrot. These options are
consistent with the findings of state-of-the-art galaxy simulations
which predict that the gas in compact SFGs reside in rotating disks
\citep{zolotov14}. In the simulations, stars have \sigstar~up to
1.5$\times$ larger depending on the projection of the gas disk. A
clear prediction of these models is that the compact quiescent
descendants should retain some rotation from its disky progenitors.

GDN-8231 stresses the need for larger samples of compact SFGs with
emission and absorption line kinematics to quantify the effects
predicted in the simulations. Those samples would allow us to study
the dependence of \ratio~with the viewing-angle and star-formation
activity. Similarly, high spatial resolution imaging in the
sub-millimeter with ALMA, or in the NIR with adaptive optics, can
provide direct measurements of the size and location of the
star-forming regions and, in some cases, resolved kinematics for the
ionized gas.

%%%%%%%DATA TABLE
\input{etab1}

\section*{Acknowledgments}

Support for Program number HST-GO-12060 was provided by NASA through a
grant from the Space Telescope Science Institute, which is operated by
the Association of Universities for Research in Astronomy,
Incorporated, under NASA contract NAS5-26555. GB acknowledges support
from NSF grant AST-08-08133. PGP-G and MCEM acknowledge support from
grant AYA2012-31277. JRT acknowledges support from NASA through Hubble
Fellowship grant \#51330. DC acknowledges support from
AYA2012-32295. The simulations were performed at NASA Advanced
Supercomputing (NAS) at NASA Ames Research Center. This work has made
use of the Rainbow Cosmological Surveys Database, which is operated by
the Universidad Complutense de Madrid (UCM), partnered with the
University of California Observatories at Santa Cruz
(UCO/Lick,UCSC). The authors recognize and acknowledge the very
significant cultural role and reverence that the summit of Mauna Kea
has always had within the indigenous Hawaiian community. We are most
fortunate to have the opportunity to conduct observations from this
mountain.  Based partly on observations made with the Gran Telescopio
Canarias (GTC), installed at the Spanish Observatorio del Roque de los
Muchachos of the Instituto de Astrofísica de Canarias, in the island
of La Palma.

\bibliographystyle{aa}
\bibliography{referencias}
\clearpage
\label{lastpage}

\end{document}

%% file: etab1.tex
%Table with properties.
%\placetable{redshifts}
%\clearpage
%\LongTables
%\begin{landscape}
\begin{deluxetable*}{cccccccccccccc}
%\tablecolumns{8}
%\rotate
%\rotate{90}
\setlength{\tabcolsep}{0.002in} 
\tablewidth{0pt}
\tabletypesize{\scriptsize}
\tablecaption{\label{redshifts} Stellar and spectroscopic properties of the MOSFIRE sample}
\tablehead{
\colhead{ID}  & \colhead{R.A.} & \colhead{DEC} & \colhead{$z_{\mathrm{spec}}$} & \colhead{F160W} &
\colhead{log~M$_{\star}$} & \colhead{r$_{e}$}  & \colhead{$n$} &
\colhead{$\sigma_{\star}$} & \colhead{$\sigma_{\rm{gas}}$}&  
\colhead{EW(\Ha)} & \colhead{\Hda}   &  \colhead{D4000}  &  \colhead{Age}\\
\colhead{(1)} & \colhead{(2)}  & \colhead{(3)} & \colhead{(4)}   & \colhead{(5)}   & 
\colhead{(6)}  & \colhead{(7)}  & \colhead{(8)}   & \colhead{(9)}  &
\colhead{(10)} & \colhead{(11)} & \colhead{(12)} &
\colhead{(13)} & \colhead{(14)} }
\startdata
8231&         189.0656 & 62.1987 & 1.674 & 21.40 & 10.75 & 2.48 & 3.95 & 215$\pm$35 & 127$\pm$32  &  5.9$^{+1.1}_{-0.7}$ &  3.6$^{+0.5}_{-0.4}$   & 1.45$\pm$0.02  & 0.75$\pm$0.12 \\
12632$^{a}$&  188.9625 & 62.2286 & 1.598 & 21.25 & 11.02 & 1.32 & 8.00 &  177$\pm$34 & -          & -0.4$^{+0.6}_{-0.8}$ &   4.3$^{+1.1}_{-1.0}$   & 1.59$\pm$0.03  & 1.07$\pm$0.20 \\
17360&        189.1153 & 62.2594 & 1.674 & 21.75 & 10.78 & 1.12 & 3.31 &  146$\pm$29 & -          & -0.7$^{+0.5}_{-1.1}$ &  3.0$^{+1.2}_{-2.0}$   & 1.62$\pm$0.03  & 1.14$\pm$0.15  \\
2617&         189.1003 & 62.1532 & 1.675 & 22.07 & 10.61 & 0.87 & 2.04 &  295$\pm$110 & -          &  0.0$^{+0.5}_{-1.5}$ &  2.6$^{+0.7}_{-1.4}$   & 1.65$\pm$0.03  &  1.31$\pm$0.27          
\enddata                                                                    
\tablecomments{\\ 
(a) Observed in \citet{newman10, belli14a}, LRIS optical spectroscopy with $\sigma=174$km~s$^{-1}$.\\
(1) General ID in the CANDELS $H$-band selected catalog in GOODS-N (Barro et al. in prep.) catalog.\\
(2,3) R.A and Declination J2000.\\
(4) Spectroscopic redshift.\\
(5) $H$ band (F160W) magnitude.\\
(6) Stellar mass  determined from SED fitting using \citet{bc03} and a \citet{chabrier} IMF, see \S~\ref{data}.\\
(7) Effective (half-light) radius (kpc) measured with GALFIT, see \S~\ref{data}.\\
(8) S\'ersic index measured with GALFIT, see \S~\ref{data}.\\
(9) Integrated velocity dispersion (km~s$^{-1}$) of the stars measured with PPXF \citep{ppxf}, see \S~\ref{kinmes}.\\
(10) Integrated velocity dispersion (km~s$^{-1}$) of the gas measured from the emission line width (FWHM), see \S~\ref{kinmes}.\\
(11) Equivalent Width of \Ha~(\AA) measured in the $H$ band MOSFIRE spectra, see \S~\ref{indexmeas}.\\
(12) Spectral index \Hda~(\AA) measured in the $Y$ band MOSFIRE spectra with PPXF, see \S~\ref{indexmeas}.\\
(13) D4000 index measured from the SED including broad- and medium- band photometry as well as {\it HST} grism spectroscopy, see \S~\ref{indexmeas}.\\
(14) Luminosity-weighted stellar age (Gyr) estimated from SED-fitting using the models of \citet{paci12}.}                                                                         
\end{deluxetable*}